# Valuing FtD Contract under Copula Approach via Monte-Carlo Stimulation

Yiran SHENG[*]

*Department of Finance, School of Economics and Management,*

*Tsinghua University, Beijing, P. R. China*

**Abstract:** This article aims to discuss some basics in field of credit modeling, specifically the pricing issue of FtD contract. We demonstrate how the popular copula approach is used in pricing FtD contract, and give a stimulation example of such practice based on SAS 9.1.

**Keyword:** first-to-default credit swap Gaussian Copula Monte-Carlo Stimulaion

---

[*] Yiran SHENG, shengyr.06@sem.tsinghua.edu.cn http://learn.tsinghua.edu.cn:8080/2006012400/index   student ID 2006012400.

# Contents

# I. Introduction

Credit derivatives were among the most popular financial products in derivative markets during the last twenty years. They are created to deal with the most critical dimension of financial risk- credit risk. Generally, their payments are exercised upon the default event happens at a unpredictable future time. Among many default triggered derivatives, default swaps (written on a single bond) and default baskets (written on a group of risky bonds) are most widely traded. Many default swaps contract have long been standardized in the OTC market or by the broker-dealer institutions who offer bid-ask two way quotes.

Among default basket swaps, the simplest and most popular one is the FtD or first-to-default contract. This article focuses on the pricing issues related to such contracts; adopting the most popular approach-both in academic sphere and in the industry practice- the copula approach via stimulations.

# II. Description of FtD contract

The seller provides protection against a basket of risky bonds. These contracts, like all swaps, have to "legs", the premium leg and the spread leg. The premium leg contains a stream of payments, offered by the buyer to the seller, until a pre-specified event, i.e. one of the bonds in the basket defaults within the life span of the FtD contract, occurs. The protection leg contains a lump sum payment to the buyer upon default, and nothing otherwise. In the event of default, the buyer will cease making spread payments and will deliver the defaulted bond to the seller. In return, the buyer receives from the seller the bond's principal and any accrued interest. If the defaulted bond has a recovery value, then on the default date, the net payment received from the seller will be the principal plus any accrued interest minus the recovery value.

The cash flow of the seller can be characterized into the following picture:

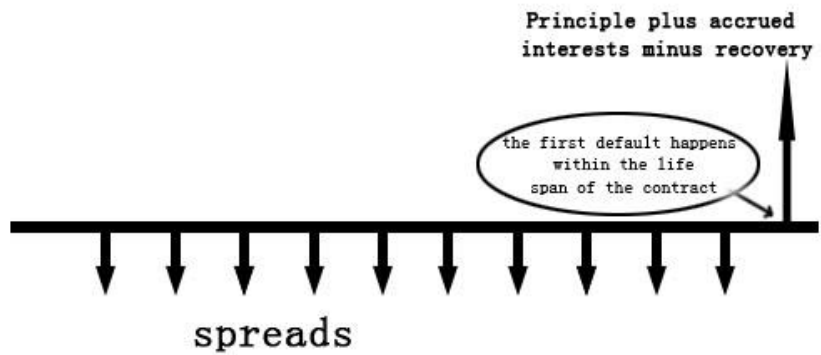

The key in pricing FtD contract is to determine the spread amount s. For simplicity, we assume there is no counter party risk involved, that is to say the buyer of FtD will not default itself before any of the bonds within the basket. As argued by (Sopranzetti, 2003), under this circumstance swaps are generally regarded as the perfect hedges for their corresponding floaters. We will carry out this assumption through this article.

## III. Modeling Default: Survive Time

Traditionally, default events are their correlation is modeled based on discrete events. (Li, 2000) was among the first to point out the many defects of this approach, and he proposed a better framework which was widely accepted in following literatures.

He introduces a random variable called the time-until-default, or simply survival time, for a security. This rv is the basis for the valuation of cash flows subject to default. To precisely determine time-until-default, we need: an unambiguously defined time origin, a time scale for measuring the passage of time, and a clear definition of default. Usually, we choose the current time as the time origin; the time scale is defined in terms of years for continuous models, or number of periods for discrete models. The meaning of default is defined by some rating agencies, such as Moody's, following the lines of (Merton, 1976).

Assume there is a security A, let continuous rv $T_A$ denote this security's time-until-default, which measures the length of time from today to the time when default occurs. For simplicity we will exchange $T_A$ with T if there is only one security. The distribution function for T is F(t):

$$F(t) = \Pr(T \leq t) \quad (1)$$

Set

$$S(t) = 1 - F(t) = \Pr(T \geq t) \quad (2)$$

S(t) is called survive function. And the pdf for T is:

$$f(t) = F'(t) = -S'(t) \quad (3)$$

In addition, we introduce another important function named hazard rate, defined as:

$$h(x) = \frac{f(x)}{1 - F(x)} = -\frac{S'(x)}{S(x)} \quad (4)$$

The hazard rate describes the instantaneous default probability for a security that has attained age x:

$$\Pr[x < T < x + \Delta x \mid T > x] = \frac{F(x + \Delta x) - F(x)}{1 - F(x)} = \frac{f(x)\Delta x}{1 - F(x)} \quad (5)$$

Then the survive function can be expressed in terms of h(x):

$$S(t) = e^{-\int_0^t h(s)ds} \quad (6)$$

Modeling hazard rate function has many advantages, and good implication to reality. First, it provides us information on the immediate default risk of each entity known to be alive at exact age t. Second, the comparisons of groups of individuals are most incisively made via the hazard rate function. Third, the hazard rate function based model can be easily adapted to more complicated situations, such as where there is censoring or there are several types of default or where we would like to consider stochastic default fluctuations.[1]

The distribution of T can be derived from historical data from rating agencies, or modeled under certain assumptions. For instance, we can assume the hazard rate to be a constant for a given security, thus $S(t) = e^{-ht}$.

## IV. Modeling FtD under Time-until-default Framework

Some basic assumptions and notations:

Let j=1 … n be the underlying securities in the FtD basket, each has a survive time $T_j$, as defined in the last section. Upon default event occurrence, each security has a recovery rate of $R_j$. Furthermore, we assume given default distribution $F_j(t)$. Denote the basket's default time as T: = min(t₁, t₂, … tₙ). For t > 0 let $B_t$ denote today's default-free zero bond price with maturity t, in other words $B_t$ is the risk-free discount factor. Let s be the spread of the contract's stream payments. We also specify the spread payments date as $t_k$ (k=1 … K), which is usually quarterly time steps. The face value of all securities is $1.

---

[1] As argued by (Li, 2000)

With risk-neutral entity, the FtD contract should be of zero-NPV, as are all financial contracts. That is the expected cash flow of the seller should be zero.

The PV of expected cash inflow (premium leg) is:

$$\sum_{k=1}^{K} s B_{t_k} P(T > t_k)$$

Note this is very similar to the equation calculating a riskless bond price, only the discount factor is not $B_{t_k}$, rather $B_{t_k} P(T>t_k)$. (Sopranzetti, 2003) is the first to express this idea, and its appliance to swap pricing.

The PV of expected cash outflow (protection leg) is, where M is the maturity of the swap:

$$\sum_{j=1}^{n} (1 - R_j) \int_0^M B_u P(T \in du, T = T_j)$$

Equate the two legs, to derive s:

$$\sum_{k=1}^{K} s B_{t_k} P(T > t_k) = \sum_{j=1}^{n} (1 - R_j) \int_0^M B_u P(T \in du, T = T_j) \quad (7)$$

Clearly, we need the probability distribution of T to determine the value of s. To do so, the joint distribution of (T₁, … Tₙ) is required. Since we already know the marginal distributions of (T₁, … Tₙ), we need a instrument to connect them to the joint distribution. There are many approaches in statistics can achieve such a task, and a copula function is the simplest and most widely used one. In this case, we adopt the simplest form of copula function, the Gaussian Copula.

# V.  Definition and Basic Properties of Copula Functions

A copula function is a function that links univariate marginals to their full multivariate distribution. For example, given m uniform random variables, U₁ … Uₘ the joint distribution function C, defined as:

$$C(u_1, ... u_m) = \Pr[U_1 < u_1, ... U_m < u_m]$$

is a copula function.

For any given multivariate distribution, F(x₁, … xₘ) and $F_j(x_j)$ j= 1 … m, the function

$$C(F_1(x_1),...F_m(x_m)) = F(x_1,...x_m)$$

Is defined using a copula function.

(Sklar, 1973) proves the inverse. He showed that any multivariate distribution function F can be written in the form of a copula function. He proved the following: If $F(x_1,… x_m)$ is a joint multivariate distribution function with univariate marginal distribution functions $F_1(x_1)$, $F_2(x_2)$, … , $F_m(x_m)$, then there exists a copula function $C(u_1, u_2, … , u_m)$ such that $F(x_1,… x_m) = C(F_1(x_1), F_2(x_2), … , F_m(x_m))$.

If each $F_i$ is continuous then C is unique. This is called the Sklar thermo. Thus, copula functions provide a unifying and flexible way to study multivariate distributions. For further information about copula functions, please refer to (张世英, 2008).

In studying the n securities, instead of making assumptions about the form of the joint survive time distribution, we can simply choose a proper copula function to describe the correlation structure between them.

# VI. FtD contract under Gaussian Copula

We assume that the joint distribution of the default times $T_1 … T_n$ is driven by a Normal copula (Gaussian copula). A Gaussian copula is the copula function that connects the marginals of multivariate normal distribution to its joint distribution. Defined as:

$$C(.) = \phi_n(\phi^{-1}(.)...\phi^{-1}(.))$$

Using C to obtain the joint distribution of $T_1 ... T_n$ :

$$F(T_1,...T_n) = C(F_1(T_1), F_2(T_2)...F_n(T_n)) \text{ or}$$

$$F(T_1,...T_n) = \phi_n(\phi^{-1}(F_1(T_1)),...,\phi^{-1}(F_n(T_n))) \quad (8)$$

Where $\phi_n$ is the n dimensional normal cumulative distribution function with correlation coefficient matrix $\Sigma$.

To simulate correlated survival times we introduce another series of random variables Y1, Y2, … Yn, such that:

$$Y_j = \phi^{-1}(F_j(T_j)) \quad (9)$$

Then there is a one-to-one mapping between Y and T . Simulating $\{T_i \mid i = 1, 2, ..., n\}$ is equivalent to simulating $\{Y_i \mid i = 1, 2, ..., n\}$. (Li, 2000) shows that the correlation between Y's is that of the underlying credits under Creditmetrics. Using CreditManager from RiskMetrics Group we can obtain the asset correlation matrix $\Sigma$.

## VII. Another Way to View Copula Approach: Default Model in Creditmetrics

(Li, 2000) is the first to point out the Creditmetrics system used copula approach implicitly, and the two frameworks are equivalent. Therefore, adopting data from Creditmetrics in Gaussian Copula Stimulation is reasonable.

The Creditmetrics framework is a kind of latent variable model in nature. It assume that default of firm i is triggered by a latent variable $X_i$. When $X_i$ fell below a certain threshold, the default scenario happens. [2]

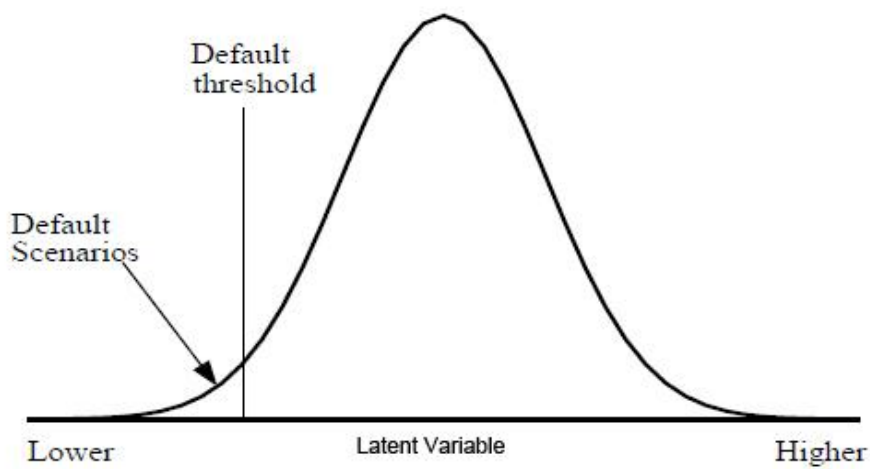

Both Creditmetrics and KMV assume $X_i$ is a linear function of risk factors $\Theta$ and idiosyncratic effects $\varepsilon_i$.

$$X_i = \sum_{j=1}^{p} \omega_{i,j} \theta_j + \sigma_i \varepsilon_i \quad (10)$$

The P dimensional vector $\Theta$ is assumed to be $N_P(0,\Omega)$; and $\varepsilon_i$'s are assumed to be iid. N(0,1). The Gaussian Structure for latent variable is equivalent as a Gaussian Copula approach, shown as follow. [3]

---

[2] This is along the lines of (Merton, 1976), where the latent variable is simply the value of the firm.
[3] For more detailed explanation please see (J.P. Morgan Creditmetrics Group., 1997)

If there is security A and B, using Creditmetrics to calculate their one-year default probability $q_A$ and $q_B$. Assume both $q_A$ and $q_B$ follows a standard normal distribution.

$$q_A = \Pr[Z_A < z_A] \quad (11)$$

$$q_B = \Pr[Z_B < z_B]$$

The joint default probability of A and B is:

$$q_{AB} = \Pr[Z_A < z_A, Z_B < z_B] = \phi_2(Z_A, Z_B, \rho) \quad (12)$$

This is equivalent to:

$$C(u = q_A, v = q_B, \rho = \gamma) = \phi_2(\phi^{-1}(u), \phi^{-1}(v), \gamma) \quad (13)$$

In addition we notice that:

$$q_i = \Pr[T_i \leq 1] = F_i(1) \quad (14)$$

$$F(1,1) = \Pr(T_A < 1, T_B < 1) = C(F_A(1), F_B(1)) \quad (15)$$

Thus we can conclude Creditmetrics uses a bivariate normal copula function and the assets correlation as the correlation. Therefore in our following stimulation, it is safe and reasonable to use data from Creditmetrics.

## VIII. A Sample Monte-Carlo Stimulation of FtD Contract

We follow the definition and notation in section VI, and make some further assumptions about input parameters. In addition, we assume $T_j$'s has constant hazard rate. Or $S_j(T)$ has a distribution described in equation (6).

| Parameters | Value |
|---|---|
| **Maturity** | M = 5 years |
| **Spread payments frequency** | $t_k$ = 0.5, 1, … 5 quarterly |
| **Discount factor** | $B_t = e^{-0.05t}$ |
| **Number of underlying securities** | n = 5 |
| **Recovery rate** | $R_j$ = 0.2 for all j's |
| **Correlation between any two assets** | $\rho$ = 10% |
| Hazard rate | $h_j$ = 0.2 for all j's |

Derived from the stimulation, the fair spread is 0.272.

-

# IX. Summary

## References

J.P. Morgan Creditmetrics Group. (1997). Creditmetrics Technical Documents.

LiXDavid. (2000). On Default Correlation: A Copula Function Approach. RiskMetrics Group.

Merton. (1976). On the Pricing of Corporate Debt: The Risk Structure of Interest Rates. Journal of Finance.

SklarA. (1973). Random Variables, Joint Distribution Functions and Copulas. Kybernetika.

SopranzettiChen and Ben J.Ren-Raw. (2003). The Valuation of Default-Triggered Derivatives. Journal of Finance and Quantative Analysis.

张世英韦艳华. (2008). Copula 理论及其在金融分析上的应用.

# Appendix

**Programming Code**

```
%let NN=10000;
proc iml;
cov={1 .1 .1 .1 .1,
     .1 1 .1 .1 .1,
     .1 .1 1 .1 .1,
     .1 .1 .1 1 .1,
     .1 .1 .1 .1 1
    };
M=5;
h={[5].2};
Rcvr={[5].2};
T={[&NN]0};
VL={[&NN]0};
VR={[&NN]0};
s={[&NN]0};
r=0.05;
rv=rannor(repeat(1200,&NN,5));
Y=rv*(root(cov));
spread=0;
i=1;
j=1;
do while (i<=&NN);
    do while (j<=5);
        x=cdf('Normal',Y[i,j]);
        Y[i,j]= -(log(1-x))/h[j];
    j=j+1;
    end;
```

```
        T[i]=min(Y[i,]);
        j=1;
        k=0.5;
        do while (k<T[i] & k<=M);
            VL[i]=VL[i]+EXP(-r*k);
            k=k+0.5;
        end;
        do while(j<=5);
            if (T[i]=Y[i,j] & T[i]<M)    then VR[i]=VR[i]+(1-Rcvr[j])*(EXP(-r*T[i]));
            j=j+1;
        end;
        j=1;

if VL[i]^=0 then s[i]=VR[i]/VL[i];
spread=spread+s[i];
i=i+1;
end;
spread=spread/&NN;
print spread;
run;
quit;
```